\documentclass[%
 aip,
 jmp,%
 amsmath,amssymb,
 reprint,%
 nofootinbib,
]{revtex4-1}
\usepackage{natbib}
\usepackage{hyperref}
\usepackage{multirow}
\usepackage{amsmath}
\usepackage{graphicx}
\usepackage{textcomp}
\usepackage{amsthm}
\usepackage{xcolor}
\usepackage{dcolumn}
\usepackage{bm}
\usepackage{algorithm}
\usepackage{algpseudocode}

\algnewcommand{\LineComment}[1]{\State \(\triangleright\) #1}
\algnewcommand{\Continue}[0]{\State \textbf{continue}}
\let\oldReturn\Return
\renewcommand{\Return}{\State\oldReturn}
\makeatletter
\let\OldStatex\Statex
\renewcommand{\Statex}[1][3]{%
  \setlength\@tempdima{\algorithmicindent}%
  \OldStatex\hskip\dimexpr#1\@tempdima\relax}
\makeatother
\algrenewcommand\algorithmicindent{1.0em}

\begin{document}
\title{Exact rule-based stochastic simulations for the system with unlimited number of molecular species}
\author{Anton V. Bernatskiy}
\email{abernats@uvm.edu}
\affiliation{Center for Complex Systems, University of Vermont, 210 Colchester Ave, Burlington, VT, 05405}
\author{Elizaveta A. Guseva}
\email{eguseva@tutanota.com}
\affiliation{Laufer Center for Physical and Quantitative Biology, Stony Brook University, Stony Brook, NY 17794-5252}

\date{\today}

\begin{abstract}
We introduce expandable partial propensity direct method (EPDM) -- a new exact stochastic simulation algorithm suitable for systems involving many interacting molecular species. 
The algorithm is especially efficient for sparsely populated systems, where the number of species that may potentially be generated is much greater than the number of species actually present in the system at any given time.
The number of operations per reaction scales linearly with the number of species, but only those which have one or more molecules.
To achieve this kind of performance we are employing a data structure which allows to add and remove species and their interactions on the fly.
When a new specie is added, its interactions with every other specie are generated dynamically by a set of user-defined rules.
By removing the records involving the species with zero molecules, we keep the number of species as low as possible.
This enables simulations of systems for which listing all species is not practical.
The algorithm is based on partial propensities direct method (PDM) by Ramaswamy et al. for sampling trajectories of the chemical master equation.
\end{abstract}
\maketitle

\begin{center}
\textbf{Frequently used abbreviations and symbols}
\end{center}
\noindent SSA -- Stochastic Simulation Algorithm\\
DM -- Direct Method\cite{Gillespie1977}\\
PDM -- Partial propensity Direct Method\cite{Ramaswamy2009}\\
SPDM -- Sorting Partial propensity Direct Method\cite{Ramaswamy2009}\\
EPDM -- Expandable Partial propensity Direct Method\\
$N$ -- number of molecular or agent species\\
$M$ -- number of all possible reactions or agent interactions\\
$\alpha$ -- maximum number of possible reactions or agent interactions between any pair of molecular or agent species


\section{Introduction}
\label{sec:intro}

Mathematical modeling of chemical reactions is an important part of systems biology research.

The traditional approach to modeling chemical systems is based on the law of mass of action equations which assumes reactions to be macroscopic, continuous and deterministic.
It doesn't account for the discreteness of the reactions or temporal and spatial fluctuations, describing only the average properties instead.
This makes it ill-suited for modeling nonlinear systems or systems with a small number of participating molecules, such as living cells.

In contrast, \textit{stochastic simulation algorithms} (SSAs) do account for discreteness and inevitable randomness of the process.
As a result, these methods are becoming increasingly popular in modern theoretical cell biology\cite{Wilkinson2011,Arkin1998,Karlebach2008}.
Additionally, they have more physical rigor compared to the empirical law of mass action ordinary differential equations\cite{Gillespie1992}.

It is worth mentioning that replacing molecules with any other interacting agents does not invalidate the approach, as long as certain conditions are met.
This makes it suitable for modeling many non-chemical systems in areas such as population ecology \cite{Reichenbach2007,McKane2005}, evolution theory \cite{Dieckmann1996,LeGalliard2005}, immunology\cite{Caravagna2010}, epidemiology \cite{Legrand2007,Breban2009,Keeling2008}, sociology \cite{Shkarayev2013,Carbone2015}, game theory \cite{Mobilia2012}, economics \cite{Juanico2012,Montero2008}, robotics \cite{Berman2009,Berman2007} and information technology \cite{Hillston2005}.

The majority of contemporary SSAs are based on the algorithm by Gillespie\cite{Gillespie1976,Gillespie1977,Gillespie1992} known as \textit{direct method} (DM).
It describes a system with a finite state space undergoing a continuous-time Markov process.
The time spent in every state is distributed exponentially; future behavior of the system depends only on its current state.
Gillespie has shown\cite{Gillespie1992} that for chemical systems these assumptions hold if the system is well-stirred and molecular velocities follow Maxwell-Boltzmann distribution.
Because of DM's good physical basis its solutions are as accurate or more accurate than the law of mass equations, which isn't necessarily a good predictor of mean values of molecular populations\cite{Gillespie1977}.

Time complexity of DM is linear in the number of reactions. 
For highly coupled systems that means that the run-time grows as a square of the number of species. 
This poses a problem for many models in systems biology which describe systems with a multitude of molecular species connected by complex interaction networks. 
Storage complexity of the DM is linear in the number of possible reactions.

Many alternative SSAs which improve the time complexity of DM (see Table \ref{tb:algorithms} for an incomplete list) were developed.
Approximate SSAs make such improvements at the cost of introducing additional approximately satisfied assumptions, while exact SSAs achieve better run-time by performing a faster computation that is equivalent to the one performed by DM.
Exact methods have been developed with time per reaction that is linear\cite{Ramaswamy2009} or, for sparsely connected networks, even constant\cite{Ramaswamy2010} in the number of molecular species.

One deficiency shared by most SSAs is that they operate with a static list of all possible reactions and species, while for some systems of interest maintaining such a list is not possible.
An example of such system is heterogeneous polymers undergoing random polymerization, an object of interest for the researchers of prebiotic polymerization and early evolutionary processes.
The number of heteropolymers that can be produced by such a process is infinite.
Even if we only consider species of length up to $L$, we have to deal with $\mathcal{O}(p^L)$ species, where $p$ is the number of monomer species.
This causes both time and storage complexity of stochastic models to grow exponentially with $L$, which limits the studies to very short polymers.

However, if an algorithm can maintain a dynamic list of molecular species and interactions, the complexity of modeling such systems can be substantially reduced.
In our example, the set of possible species is so large that even for moderate cutoff $L$ the vast majority of all possible species will have a population of zero.
If at any time step a specie has a population of zero, no reaction can happen in which this specie is a reagent.
Therefore, the reactions involving such a specie need not be tracked until some reaction occurs in which the specie is a product.

In addition, many studies are concerned with emergent phenomena, which often involve the system exhibiting some dynamical pattern, e.g. sitting in a stable or chaotic attractor.
Such phenomena tend to only involve a subset of possible states and transitions and not explore the space of all possibilities uniformly.
This may limit the species and reactions involved in the phenomenon to a small subset of all possible species and reactions.
For example, in the study by Guseva et al.\cite{Guseva2016} effective populations of all sequences are $\propto 10^3$, while the number of all the possible species is $\propto 10^7$.
The model is tightly coupled, thus time costs are $10^4$ times higher and memory costs are $10^8$ times higher than they could be if only the final subset of reactions and species was considered and the SSA retained the time and storage complexity of the best SSAs for static lists (e.g. PDM\cite{Ramaswamy2009}). 

However, not only this subset is unknown prior to the experiment, but the transient to the behavior of interest may involve many more species and reactions than the behavior itself.
Thus, the simulation cannot be constrained to use a smaller list of species and reactions if such list is static.

One class of systems in which reactions cannot be listed occurs in solid state chemistry.
To simulate those, Henkelman and Jonsson\cite{Henkelman2001} developed an algorithm in which the list of possible reactions is generated on the fly and used in the standard Gillespie's algorithm.
This approach was formulated as a part of simulation framework specific to solid state chemistry.
It remained obscure outside of the solid state chemistry community and was rediscovered independently by authors of the present work early in the course of its preparation.

Here we present \textit{extendable partial-propensity direct method} (EPDM): a general purpose, exact SSA in which species and reactions can be added and removed on the fly.
Unlike Henkelman and Jonsson's approach, our algorithm is based on the \textit{partial-propensity direct method} (PDM) by Ramaswamy et al.\cite{Ramaswamy2009} and retains its linear time complexity in the number of species. 
Storage complexity is linear requirements in the number of reactions.

Only the species with nonzero populations and the reactions involving them are recorded and factored into the complexity.
This reduces time complexity of executing one reaction by a factor of $N_{tot}/N$ where $N_{tot}$ is the number of all possible species and $N$ is the number of species with nonzero population at the time of the reaction.
Storage complexity is reduced by the square of that factor for densely connected reaction networks.

We test the performance of a C++ implementation of our algorithm\footnote{https://github.com/abernatskiy/epdm} for two chemical systems, investigate scaling and compare the performance with two other SSAs.
The results confirm our predictions regarding the algorithm's complexity.

\begin{table} [h]
\begin{tabular}{| p{0.7in} |p{2.6in} |}
\hline
  Exact & 
  \textbullet\,Direct\,method\,(DM) \cite{Gillespie1976,Gillespie1977}\newline
\textbullet\,First\,reaction\,method\,(FRM)\,\cite{Gillespie1976}\newline
 \textbullet\,Gibson-Bruick's\,next-reaction\,method (NRM)\cite{Gibson2000}\newline
 \textbullet\,Optimized direct method (ODM)~\cite{Cao2004}\newline
 \textbullet\,Sorting direct method (SDM)~\cite{McCollum2006}\newline
 \textbullet\,Partial propensity direct method (PDM)~\cite{Ramaswamy2009}
  \\ \hline 
  Approximate &
  \textbullet\,$\tau$-leaping~\cite{Gillespie2001,Cao2005,Cao2006,Peng2007}\newline
 \textbullet\,$k_\alpha$-leaping~\cite{Gillespie2001}\newline
 \textbullet\,Implicit $\tau$-leaping~\cite{Rathinam2003}\newline
  \textbullet\,The slow-scale method~\cite{Cao2005a} \newline
 \textbullet\,$R$-leaping~\cite{Auger2006} \newline
  \textbullet\,$L$-leap~\cite{Peng2007a}\newline
 \textbullet\,$K$-leap~\cite{Cai2007} 
 \\    \hline
\end{tabular}
\caption{List of some exact and approximate stochastic simulation algorithms}
\label{tb:algorithms}
\end{table}

\section{Direct stochastic algorithms now}

\subsection{Gillespie Algorithm}
\label{sec:gillespie}

Being an exact SSA, our algorithm performs an optimized version of the same basic computation as the Gillespie's DM\cite{Doob1953,Gillespie1977}.
DM is based on the following observation:
\begin{quote}
	Probability that any particular interaction of agents occurs within a
    small period of time is determined by and 
    proportional to a product of a rate constant specific to this interaction and 
    a number of distinct combinations of agents required for the interaction.
\end{quote}
Since we developed EPDM with chemical applications in minds, we will hereafter refer to agents of all kinds as molecules.
However, as long as the observation above holds, the method is valid for any other kind of agent (see Introduction).

The algorithm starts with initialization (\textbf{Step 1}) of the types and numbers of all the molecules initially present in the system, reaction rates and the random number generator.
Then propensities of all reactions are computed (\textbf{Step 2}). Propensity $a_\mu$ of a reaction $R_\mu$ ($\mu\in\{1,2,...,M\}$) is proportional to its reaction rate $c_\mu$:
\begin{equation}
\label{eq:gillespiePropensity}
	a_\mu=h_\mu c_\mu.
\end{equation}
Here, $h_\mu$ is the number of distinct molecular reactant combinations for reaction $R_\mu$ at the current time step, a combinatorial function which depends on the reaction type and the numbers of molecules of all reactant types 
\cite{Gillespie1976}. Total propensity is the sum of propensities of all reactions:
\begin{equation}
\label{eq:gillespieTotalPropensity}
	a=\sum_{\mu=1}^M a_\mu.
\end{equation}

The next step (\textbf{Step 3}), called Monte Carlo step or sampling step, is the source of stochasticity. 
Two real-valued, uniformly distributed random numbers from $[0,1)$ are generated. 
The first ($r_1$) is used to compute time to next reaction $\tau$:
\begin{equation}
\label{eq:tauGillespie}
	\tau = \frac{1}{a}\ln\left( \frac{1}{r_1}  \right).
\end{equation}
The second one ($r_2$) determines which reaction occurs during the next time step $\tau$. 
The $j$-th reaction occurs if
\begin{equation}
\label{eq:reactionGillespie}
	\sum_{\mu=1}^{j-1}a_j \leqslant a r_2 < \sum_{\mu=1}^j a_j.
\end{equation}

The next step is update (\textbf{Step 4}): simulation time is increased by $\tau$ generated at Step 3,
molecules counts are updated using the stoichiometric numbers of the sampled reaction and 
propensities are updated in accordance with the new molecular counts. 

The last step is iteration: go back to Step 3 unless some termination condition is met.
Termination should occur if no further reactions are possible (i.e. when the total propensity $a=0$).
Optional termination conditions may include reaching a certain simulation time, performing a given number of reactions, reaching some steady state etc.

At every step the algorithm looks through the list of all $M$ possible reactions.
Therefore, the time it takes to process one reaction (i.e. perform Steps 3 and 4) is proportional to $M$:
\begin{equation}
\label{eq:gillespieTimeComplexity}
	t_{reac} = \mathcal O(M).
\end{equation}
There must be a record for every reaction, so the space complexity is also $\mathcal O(M)$.


\subsection{Partial propensity methods}

For many systems it is valid to neglect reactions which involve more than two molecules or agents.
This premise allows for a class of partial-propensity direct methods.
The method described in the present work is among those.

If $\alpha$ is the maximum number of reactions which may happen between any pair of the reagent species then $M=\mathcal O(\alpha N^2)$.
Then the expression for the time complexity of DM \eqref{eq:gillespieTimeComplexity} becomes quadratic in $N$:
\begin{equation}
	t_{reac} = \mathcal O(\alpha N^2).
\end{equation}
Partial-propensity direct method (PDM) and sorting partial-propensity direct method (SPDM) \cite{Ramaswamy2009} improve this bound to $\mathcal O(\alpha N)$ by \emph{associating each reaction with one of the involved reagents} and \textit{sampling the reactions in two stages}.
In the first stage the first reactant specie of the reaction to occur is determined; this takes $\mathcal O(N)$ operations.
The second stage determines the second specie and a particular reaction to occur; this takes $\mathcal O(\alpha N)$ operations, and the total complexity of the sampling step adds up to $\mathcal O(\alpha N)$.

Since any specie can be involved in at most $\mathcal O(\alpha N)$ uni-molecular or bi-molecular reactions, only $\mathcal O(\alpha N)$ values have to be updated when the molecular counts change.
This enables PDM and SPDM to perform the update step without worsening the time complexity of the sampling step.


The final time complexity of these algorithms
\begin{equation}
\label{eq:pdmbound}
	t_{reac} = \mathcal O(\alpha N)
\end{equation}
holds irrespective of the degree of coupling of the reaction network.
The number of records required for sampling is still $\mathcal O(M)=\mathcal O(\alpha N^2)$.

Note that for sparsely coupled reaction networks, time complexity can be improved further to 
$\mathcal O(1)$\cite{Ramaswamy2010}; here we won't concern ourselves with such reaction networks.


\section{Algorithm description}

\subsection{Reaction grouping}
\label{sec:reactionGrouping}

Similar to PDM and SPDM\cite{Ramaswamy2009}, our method relies on splitting the reactions into groups associated with one of the reagents.
To simplify this procedure, we reformulate all reactions with up to two reagents as bimolecular by introducing the virtual \textit{void specie} $\varnothing$ which always has a molecular count of 1.
It can interact with other, real species and itself, however its stoichiometry is always zero.
With these new assumptions, unimolecular reactions are reformulated as follows:
\begin{equation}
\label{eq:reformulationUni}
\begin{aligned}
	&S_i \rightarrow \mbox{products} \Rightarrow \\
	&1\, S_i + 0\, \varnothing \rightarrow \mbox{products}.
\end{aligned}
\end{equation}
Source reactions are also reformulated:
\begin{equation}
\label{eq:reformulationSource}
\begin{aligned}
	&\varnothing \rightarrow \mbox{products} \Rightarrow\\
	&0\,\varnothing + 0\,\varnothing \rightarrow \mbox{products}.
\end{aligned}
\end{equation}

Our algorithm keeps a list of species existing in the system to which entries can be added.
When a new specie $S_i$ is added to the list, the algorithm considers every specie $S_j$ in the updated list.
For each (unordered) pairing $\{S_i, S_j\}$ a list of possible reactions is generated.
If the list is not empty, it is associated with $S_j$, the specie that has been added to the list earlier unless it coincides with $S_i$.
If $S_i=S_j$, the list is associated with $S_i$.

For example, the first step of the initialization stage involves adding the first element to the list of species which is always the void specie $\varnothing$.
At this point there are no species in the list aside from $\varnothing$, so the algorithm checks which kinds of reactions may happen between $\varnothing$ and itself.
Due to the reformulation \eqref{eq:reformulationSource} this will involve all source reactions.
Their list will be generated and associated with the newly added specie $\varnothing$.

Another example. Suppose the list of known species is $[\varnothing, S_1]$ and we're adding a specie $S_2$ which reacts with $S_1$, itself and also participates in some unimolecular reactions.
The list becomes $[\varnothing, S_1, S_2]$ and the algorithm proceeds to pair up $S_2$ with every specie in it and generate reaction lists.

Due to the reformulation \eqref{eq:reformulationUni} it will find all the unimolecular reactions for the pair $\{S_2, \varnothing\}$. The list of these reaction will be associated with the previously known specie of the pair, $\varnothing$.

When considering the pair $\{S_2, S_1\}$ it will find all the reactions between $S_1$ and $S_2$ and associate them with $S_1$.

Finally, it will consider a pair $\{S_2, S_2\}$ and find its reactions with itself. Since there are no previously known species in the pair, the reactions will be associated with $S_2$.

\subsection{Propensities generation}
\label{sec:propensitiesGeneration}

As the reactions are generated and associated with species we also compute and store some propensities.

For all reactions $R_\mu$ with up to two participating molecules full propensities $a_\mu$ are
\begin{equation}
\begin{aligned}
	a_\mu &= n_i n_j c_\mu \mbox{ for a bimolecular reaction}\\
    & \mbox{of distinct species }i\neq j,\, S_i + S_j \rightarrow \mbox{products},\\
	a_\mu &= \frac12 n_i (n_i-1) c_\mu \mbox{ for a bimolecular reaction}\\
    & \mbox{of identical species }2S_i \rightarrow \mbox{products},\\
    a_\mu &= n_i c_\mu \mbox{ for a unimolecular reaction }\\
    	&\hspace{4cm}S_i \rightarrow \mbox{products},\\
    a_\mu &= c_\mu \mbox{ for a source reaction }\varnothing \rightarrow \mbox{products}.
\end{aligned}
\end{equation}
Since the void specie $\varnothing$ has a fixed population of 1 we can use the formula for the unimolecular reactions for source reactions as well: $a_\mu = 1\cdot c\mu$ for $\varnothing \rightarrow \mbox{products}$.
Then for all reactions we can define \textit{partial propensity w.r.t. the reactant specie $S_i$} as
\begin{equation}
\label{eq:partialPropensityDefinition}
	\pi^{(i)}_\mu \equiv a_\mu / n_i.
\end{equation}
For the reactions with up to two reagents, partial propensities are\cite{Ramaswamy2009}
\begin{equation}
\label{eq:partialPropensityForReactionTypes}
\begin{aligned}
	\pi^{(i)}_\mu &= n_j c_\mu \mbox{ for a bimolecular reaction}\\
    & \mbox{of distinct species }i\neq j,\, S_i + S_j \rightarrow \mbox{products},\\
	\pi^{(i)}_\mu &= \frac12 (n_i-1) c_\mu \mbox{ for a bimolecular reaction}\\
    & \mbox{of identical species }2S_i \rightarrow \mbox{products},\\
    \pi^{(i)}_\mu &= c_\mu \mbox{ for a unimolecular reaction }\\
    	&\hspace{4cm}S_i \rightarrow \mbox{products},\\
    \pi^{(\varnothing)}_\mu &= c_\mu \mbox{ for a source reaction }\varnothing \rightarrow \mbox{products}.
\end{aligned}
\end{equation}

Suppose some specie $S_j$ is added to the list of known species as described in section \ref{sec:reactionGrouping}.
For every specie $S_i$ in the updated list we generate a list of possible reactions between $S_i$ and $S_j$ and associate it with $S_i$.
For every reaction $R_{ijk}$ in the list, we compute its rate constant $c_{ijk}$ and its partial propensity $\pi^{(i)}_{ijk}$ w.r.t. $S_i$ using formulas \eqref{eq:partialPropensityForReactionTypes}.

We also keep some sums of propensities to facilitate the sampling.
Given a list of reactions between $S_i$ and $S_j$ associated with $S_i$, we define $\Psi^{(i)}_{ij}$ as
\begin{equation}
\label{eq:psiDefinition}
	\Psi^{(i)}_{ij} = \sum_{k=1}^{\alpha_{ij}} \pi_{ijk}^{(i)},
\end{equation}
where $\alpha_{ij}$ is the number of reactions possible between $S_i$ and $S_j$.

$\Lambda^{(i)}_i$ is the sum of partial propensities of all reactions associated with $S_i$:
\begin{equation}
\label{eq:lambdaDefinition}
	\Lambda^{(i)}_i = \sum_{j=1}^{m_{i}} \Psi^{(i)}_{ij}.
\end{equation}
Here, $m_{i}$ is the number of lists of reactions with other species associated with $S_i$.

$\Sigma_i$ is the full propensity of all reactions associated with $S_i$:
\begin{equation}
\label{eq:sigmaDefinition}
	\Sigma_i = n_i \Lambda^{(i)}_i.
\end{equation}
and $a$ is defined as a total full propensity of all reactions in the system:
\begin{equation}
\label{eq:aDefinition}
	a = \sum_{i=1}^N \Sigma_i.
\end{equation}

\begin{table*}[ht]
\begin{tabular}{|p{1in}|p{1.1in}|p{1in}|p{1in}|p{0.6in}|p{1.8in}|}
\hline
\multirow{2}{*}{Data type}
	& \multicolumn{4}{c|}{Necessary members} & \multirow{2}{*}{Significance}\\\cline{2-5}
    & Containers & Scalars & References & Functions & \\
\hline
\texttt{TotalPopulation} &
Linked list of \texttt{Population}s &
Total propensity $a$ \eqref{eq:aDefinition}, current time $t$, random generator state &
-- &
-- &
Represents the whole system being modeled\\
\hline
\texttt{Population} & 
Linked list of \texttt{Relation}s, linked list of \texttt{RelationAddress}es &
\texttt{Specie} $S$, molecular count $n$, propensity sums $\Sigma$ \eqref{eq:sigmaDefinition} and $\Lambda$ \eqref{eq:lambdaDefinition} &
-- &
-- &
Represents a population of molecules of a specie $S$\\
\hline
\texttt{Relation} &
Linked list of \texttt{Reaction}s &
Partial propensity sum $\Psi$ \eqref{eq:psiDefinition} &
To \texttt{RelationAddress} pointing to this \texttt{Relation}&
-- &
Data on all reactions possible between a pair of species, to be stored withing the list of an \textit{owner} specie\\
\hline
\texttt{RelationAddress} &
-- &
-- &
To a \texttt{Relation}, its owner's \texttt{Population}, list containing this \texttt{RelationAddress} and itself &
-- &
Reference to a \texttt{Relation} to be kept by a non-owner specie, useful in propensity updates and population deletions\\
\hline
\texttt{Reaction} &
Array of pairs $(ID,\sigma)$ of $ID$s and stoichiometries of participating species &
Rate constant $c$, partial propensity $\pi$ w.r.t. the owner specie \eqref{eq:partialPropensityForReactionTypes} &
-- &
-- &
Represents a reaction\\
\hline
\texttt{Specie} &
-- &
Unique, compact specie identifier $ID$ &
-- &
\texttt{Specie:: reactions (Specie)} &
An extended specie representation capable of keeping extra information to generate lists of possible reactions with \texttt{reactions()} method quickly\\
\hline
\end{tabular}
\caption{List of data structure types used by the algorithm}
\label{tb:structures}
\end{table*}

\subsection{Data model}
\label{sec:dataStructure}

\begin{figure*}[hbt!]
\centering
\includegraphics[width=0.95\textwidth]{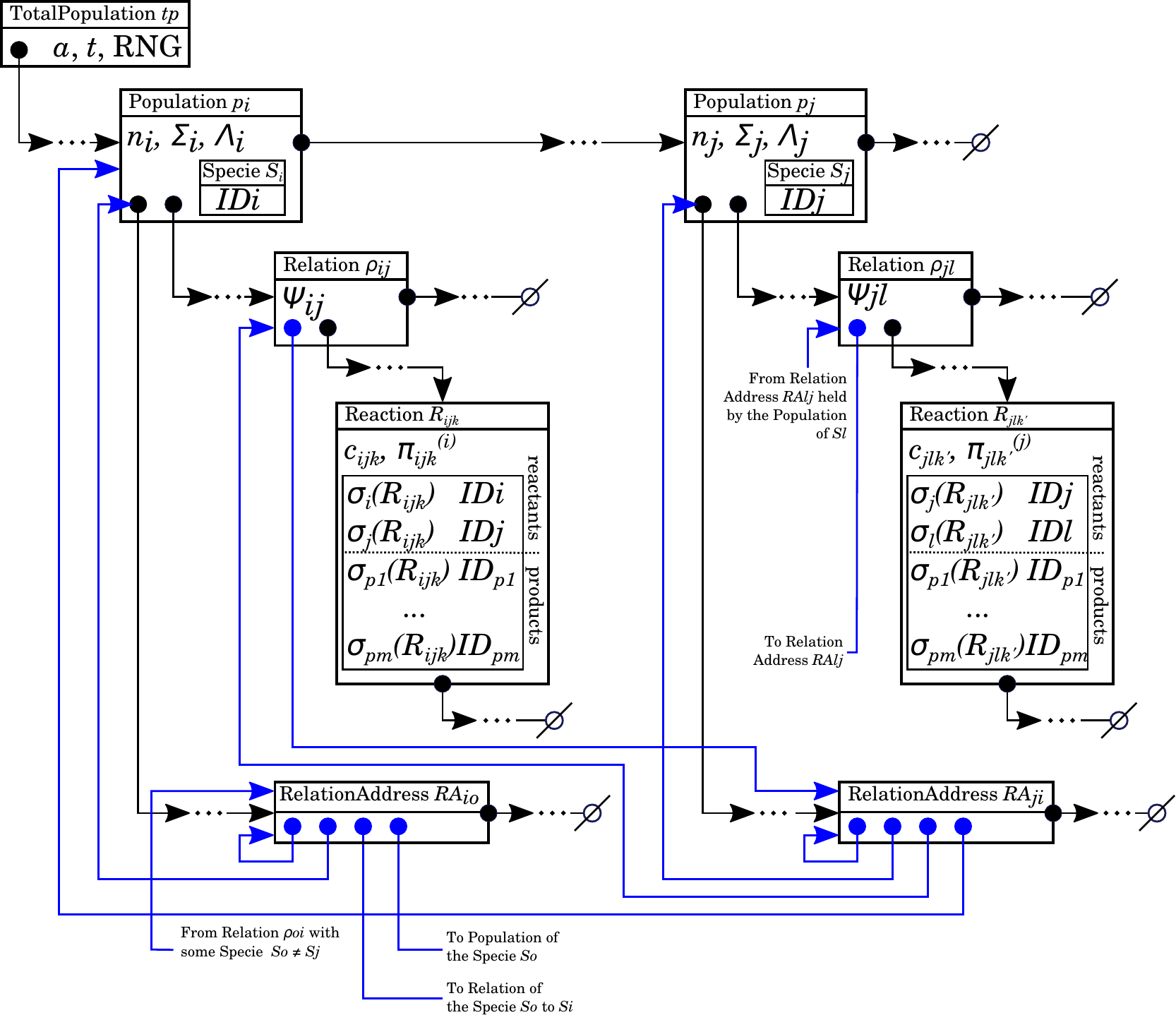}
\caption{EPDM data model. Circles with arrows denote references. References shown in black form the hierarchical linked list; in implementation, many of those are hidden within the \texttt{std::list} container. Auxiliary references added for dynamic updates are shown in blue. Letters and boxes within the boxes indicate non-reference member variables. Crossed empty circle represents a null reference.
}
\label{fig:epdm_data_structure}
\end{figure*}

Similar to Steps 3 and 4 in Gillespie's DM (see section \ref{sec:gillespie}), execution of each reaction in our algorithm involves two steps: \textit{sampling} and \textit{updating}.
We'll refer to the data used in the sampling process as \textit{primary} and call all the data which is used only for updating \textit{auxiliary}.

To minimize the overhead associated with adding and removing the data we use a hierarchical linked list (multi-level linked list of linked lists).
We define several data structure types (see table \ref{tb:structures}), many of which include lists of instances of other types.
Complete data model is shown in Figure \ref{fig:epdm_data_structure}.

Top level list is stored within a structure of type \texttt{TotalPopulation}.
Aside from the list, the structure holds the data describing the system as a whole: total propensity $a$ (see \eqref{eq:aDefinition}), current simulation time $t$ and the random generator state.

Elements of the top level list are structures of type \texttt{Population}, each of which holds the information related to the molecular population of a particular specie $S_i$.
The information about the specie itself, to which we will in the context of its \texttt{Population} refer as the \textit{owner specie}, is represented by a member object of user-defined class \texttt{Specie} (see note \ref{note:specieClass} at the end of this section), containing a unique string specie identifier $ID_i$ as a member variable. \texttt{Population} also contains the number of molecules in the population $n_i$, total propensity of the population $\Sigma_i$ and total partial propensity $\Lambda^{(i)}_i$ w.r.t. $S_i$ of all reactions associated with $S_i$ (see eqs. \eqref{eq:sigmaDefinition} and \eqref{eq:lambdaDefinition}).

The \texttt{Population} structures are appended to the top level list sequentially over the course of the algorithm execution (see section \ref{sec:reactionGrouping}).
Each of them holds, in addition to the data mentioned above, two linked lists: of structures of type \texttt{Relation} and of structures of type \texttt{RelationAddress}.
The former stores the list of all reactions possible between the owner specie and some other specie that has been added later in the course of the algorithm's operation.
\texttt{RelationAddress} structures hold the references used to access \texttt{Relation}s in which the owner specie participates, but not as an owner.
Those include all the species added before the owner.

It can be observed that the \texttt{Population} which has been added first will necessarily has an empty list of \texttt{RelationAddress}es and a potentially large list of \texttt{Relation}s, with as many elements as there are species with which the first specie has any reactions.
On the other hand, the \texttt{Population} added last will not be an owner of any \texttt{Relation}s and its list of those will be empty. In this case, all information about the specie's reactions is owned by other species and only available within the \texttt{Population} through the list of \texttt{RelationAddress}es.

A \texttt{Relation} structure owned by $S_i$ holds a linked list of all possible \texttt{Reaction}s between $S_i$ and $S_j$ and their total partial propensity $\Psi^{(i)}_{ij}$ (see eq. \eqref{eq:psiDefinition}).
\texttt{Reaction}s store the reaction rate $c_{ijk}$, partial propensity $\pi^{(i)}_{ijk}$, a table of stoichiometric coefficients and $ID$s of all participating species, including products.
Additionally, each of them stores an auxiliary reference to the \texttt{RelationAddress} structure pointing to this \texttt{Relation}.

\texttt{RelationAddress} is an auxiliary structure holding references to a \texttt{Relation} in which the owner specie participates without owning it, and to the \texttt{Population} of the specie which does own the \texttt{Relation}.
It also contains the references to itself and to the list holding it.

Notes:
\begin{enumerate}
\item We will say that a \texttt{TotalPopulation} is \textit{valid} iff all variables mentioned in eqs. (\ref{eq:partialPropensityForReactionTypes}--\ref{eq:aDefinition}) satisfy these equations and all references mentioned in the present section are valid and point where intended.
\item \label{note:specieClass} Unique specie string $ID$ format and \texttt{Specie} objects are user-defined and must be freely convertible between each other. The utility of having a separate \texttt{Specie} class lies in it having a user-defined method $\mbox{\texttt{reactions()}}: (self, \mbox{\texttt{Specie}})\rightarrow ListOfReactions$ which produces a list of all reactions possible between the specie described by a caller object and a specie described by the the argument object. For some systems, this computation can be made much faster if some auxiliary data can be kept in the structure describing a specie. String $ID$s on the other hand are intended as memory efficient representations of specie data which are used when this computation may not be needed, e.g. to represent reaction product not yet present in the system or for specie comparison.

Our method does not rely on this separation other than as a means of optimization.
\item Additionally, in our implementation we provide a global structure to hold the parameters of the system which may influence the behavior of the \texttt{reactions()} method of class \texttt{Specie}.
\end{enumerate}

\subsection{Adding a \texttt{Population}}
\label{sec:addition}

\begin{figure}[t]
\begin{algorithm}[H]
\caption{Adding a \texttt{Population}}
\label{al:addition}
\begin{algorithmic}[1]
\Function {AddPopulation} {$tp$, $ID$, $n$}
	\State Append a new \texttt{Population} $p_N$ to $tp$'s list
    \State Convert $ID$ to a \texttt{Specie} object $S$
    \State $S_N \gets S$, $n_N \gets n$
    \State $\Lambda^{(N)}_{N} \gets 0$, $\Sigma_{N} \gets 0$
	\ForAll{\texttt{Population}s $p_i$ in $tp$'s list}
    	\State $ListOfReactions \gets S_N.\mbox{\texttt{reactions}}(S_i)$
        \If{$ListOfReactions$ is empty}
        	\Continue
        \EndIf
        \State Append a new \texttt{Relation} $\rho_{iN}$ to $p_i$'s list
        \State Append all \texttt{Reaction}s in $ListOfReactions$
        \Statex[3] to $\rho_{iN}$'s list
        \State $\Psi^{(i)}_{iN} \gets 0$
        \ForAll{\texttt{Reaction}s $R_{iNk}$ in $\rho_{iN}$'s list}
        	\State Compute $\pi^{(i)}_{iNk}$ using \eqref{eq:partialPropensityForReactionTypes}
            \State $\Psi^{(i)}_{iN} \gets \Psi^{(i)}_{iN} + \pi^{(i)}_{iNk}$
       	\EndFor
        \State $\Lambda^{(i)}_i \gets \Lambda^{(i)}_i + \Psi^{(i)}_{iN}$
       	\State Append a new \texttt{RelationAddress} structure $RA_{Ni}$
        \Statex[3] to $p_N$'s list
        \State Store references to $\rho_{iN}$, $p_i$, $RA_{Ni}$ and $p_N$'s list of
        \Statex[3] \texttt{RelationAddress}es at $RA_{Ni}$
        \State Store reference to $RA_{Ni}$ at $\rho_{iN}$
    \EndFor
    \State Recompute $a$ using \eqref{eq:aDefinition}
\EndFunction
\end{algorithmic}
\end{algorithm}
\end{figure}

Suppose that we have a valid \texttt{TotalPopulation} with $N-1$ \texttt{Population}s.
To add a \texttt{Population} based on a specie $ID$ and a molecular count $n$, we follow the method described in section \ref{sec:reactionGrouping} (see also algorithm \ref{al:addition}).
We begin by appending a \texttt{Population} $p_N$ to \texttt{TotalPopulation}'s list. $p_N$'s \texttt{Specie} structure $S_N$ is converted from $ID$, its molecular count $n_N \gets n$ and propensities $\Lambda^{(N)}_{N} \gets 0$, $\Sigma_{N} \gets 0$.

Next, for each \texttt{Population} $p_i$ in \texttt{TotalPopulation}'s updated list (including the newly added $p_N$) we compute a list of all possible reactions between $S_N$ and $p_i$'s \texttt{Specie} $S_i$.
If the list is not empty, it is then converted into \texttt{Relation} $\rho_{iN}$ and stored at $p_i$'s list of those.
To do the conversion, we compute partial propensities $\pi^{(i)}_{iNk}$ and $\Psi^{(i)}_{iN}$ using eqs. \eqref{eq:partialPropensityForReactionTypes} and \eqref{eq:psiDefinition}.
We also store copies of $S_i$'s $ID$ at every \texttt{Reaction} in the list and at the \texttt{Relation} $\rho_{iN}$ itself, to keep track of the specie w.r.t. which the current partial propensities are computed.
$\rho_{iN}$'s reference to the \texttt{RelationAddress} structure is invalid at this point.

After appending the \texttt{Relation} we update $p_i$'s propensities: $\Lambda^{(i)}_i \gets \Lambda^{(i)}_i + \Psi^{(i)}_{iN}$, $\Sigma_i \gets n_i\Lambda^{(i)}_i$.

We then proceed to create a \texttt{RelationAddress} structure $RA_{Ni}$ pointing to $\rho_{iN}$. A blank structure is appended to $p_N$'s list.
We take references to $\rho_{iN}$, $p_i$, $RA_{Ni}$ and $p_N$'s list of \texttt{RelationAddress}es and save them at $RA_{Ni}$.

Then, we take the reference to $RA_{Ni}$ and store it at $\rho_{iN}$. At this point, all references in the whole structure are valid and correct.

Processing each $p_i$ in the \texttt{TotalPopulation}'s list takes $\mathcal{O}(\alpha)$, so the whole procedure up to this point takes $\mathcal{O}(\alpha N)$.
At this point we need to update the total propensity of the system $a$ by recomputing it, which takes $\mathcal{O}(N)$\footnote{It can be done during the iteration in $\mathcal{O}(1)$, but we chose to recompute it for improved numerical accuracy.}.

The total number of operations it takes to add one \texttt{Population} is $\mathcal{O}(\alpha N)$.
The operation preserves the validity of the data structure.

\subsection{Initialization stage}
\label{sec:initialization}

To initialize the data structure, we make a \texttt{TotalPopulation} with an empty list of \texttt{Population}s, initialize $a\gets 0$, $t\gets 0$ and the random generator with a seed.
The only variable that needs to take a particular value for the \texttt{TotalPopulation} to be valid is $a$ and it has the correct value of 0; therefore, it is a valid \texttt{TotalPopulation}.
We build the data structure by adding the initial populations to this structure as described in section \ref{sec:addition}.
The resulting structure is valid because we only used validity-preserving operatins.

Adding every population takes $\mathcal{O}(\alpha N)$ operations and it must be repeated $N$ times.
This brings the complexity of the initialization step to $\mathcal{O}(\alpha N^2)$.

\subsection{Deleting a \texttt{Population}}
\label{sec:deletion}

\begin{figure}[b]
\begin{algorithm}[H]
\caption{Deleting a \texttt{Population}}
\label{al:deletion}
\begin{algorithmic}[1]
\Function {DeletePopulation} {$tp$, $p_i$}
	\ForAll{\texttt{Relation}s $\rho_{ij}$ in $p_i$'s list}
    	\State Follow $\rho_{ij}$'s reference to \texttt{RelationAddress}
        \Statex[3] pointing at it, $RA_{ji}$
        \State Use the reference at $RA_{ji}$ to itself and to the
        \Statex[3] list holding it to remove it from the list
    \EndFor
	\ForAll{\texttt{RelationAddress} $RA_{ij}$ in $p_i$'s list}
    	\State Follow $RA_{ij}$'s reference to the \texttt{Relation} $\rho_{ji}$
        \Statex[3] and to the \texttt{Population} $p_j$ owning it
        \State Remove $\rho_{ji}$ from $p_j$'s list
    \EndFor
    \State Delete all the date in the $p_i$ structure
    \State Remove $p_i$ from $tp$'s list
\EndFunction
\end{algorithmic}
\end{algorithm}
\end{figure}

Our algorithm is designed to keep track only of the species with a nonzero molecular count and reactions involving them.
To accomplish that, we use the addition operations described in the previous section and deletion operation described here (see also algorithm \ref{al:deletion}).
Deletion is only 	ever applied to \texttt{Populations} of species with zero molecules.
All propensities of reactions involving such species are zeros; this enables us to simplify the procedure.

To remove a \texttt{Population} $p_i$, we begin by removing all \texttt{RelationAddress} structures pointing at \texttt{Relation}s in $p_i$'s list.
Each \texttt{Relation} $\rho_{ij}$ in $p_i$'s list contains a reference to the \texttt{RelationAddress} structure pointing at it, $RA_{ji}$, which in turn contains a reference to itself and to a list holding it.
We use those to remove each $RA_{ji}$ from its list.
Since $S_i$ can be involved in at most $N$ relations, this step takes $\mathcal{O}(N)$ operations.

Next, we remove all \texttt{Relation}s in which $S_i$ participates, but which are owned by other species.
For each \texttt{RelationAddress} $RA_{ij}$ we follow its references to the \texttt{Population} $p_j$ of the other specie and to its \texttt{Relation} $\rho_{ji}$ with $S_i$.
We use those to remove $\rho_{ji}$ from $p_j$'s list.
This step also takes $\mathcal{O}(N)$ operations.

Finally, we delete the whole \texttt{Population} structure $p_i$ from \texttt{TotalPopulation}'s list.
We recursively remove all the structures in its lists, which takes $\mathcal{O}(N)$ operations for the list of \texttt{RelationAddress}es and $\mathcal{O}(\alpha N)$ operations for the list of \texttt{Relations}.
The resulting \texttt{TotalPopulation} is valid since all the propensities of the reacitons involving $S_i$ are zeros and the remaining propensity sums has not changed; it also contains no invalid references.

The final complexity of the deletion operation is $\mathcal{O}(\alpha N)$.

\subsection{Sampling stage}
\label{sec:sample}

Similarly to DM\cite{Gillespie1976} and PDM\cite{Ramaswamy2009}, our algorithm simulates the system by randomly sampling time to the next reaction and the reaction itself with certain distributions.
Here we describe how it happens in our algorithm.

We begin by generating two random numbers $r_1$ and $r_2$.
The first one is used to compute the time to the next reaction $\tau$ exactly as in DM and PDM (see eq. \eqref{eq:tauGillespie}). The second random number $r_2$ is used to sample the reaction similarly to how its done in PDM\cite{Ramaswamy2009}.

The sampling process has three stages (see algorithm \ref{al:reactionSampling}).
During the first stage (lines 2-6) we determine the first specie participating in the reaction to happen.
To this end, we go through the list of \texttt{Population}s $p_1 ... p_N$ until the following condition is satisfied:
\begin{equation}
\label{eq:samplingPopulation}
	\sum_{i=1}^{J} \Sigma_i \leqslant a r_2 < \sum_{i=1}^{J+1} \Sigma_i.
\end{equation}

The second stage (lines 7-11) involves finding the second reactant.
We look for a \texttt{Relation} $\rho_{JK}$ among those attached to the \texttt{Population} $p_J$ for which the following condition holds:
\begin{equation}
\label{eq:samplingRelation}
	\sum_{i=1}^{J} \Sigma_i + n_J \sum_{i=1}^K \Psi^{(J)}_{Ji} \leqslant a r_2 < 
    \sum_{i=1}^{J} \Sigma_i + n_J \sum_{i=1}^{K+1} \Psi^{(J)}_{Ji}
\end{equation}

During the third stage (lines 12-16), we determine which of the reactions possible between the two species is going to happen.
We go through $\rho_{JK}$'s list of the \texttt{Reaction}s, looking for a reaction $R_{JKL}$ such that
\begin{equation}
\label{eq:samplingReaction}
\begin{aligned}
	\sum_{i=1}^{J} \Sigma_i &+ n_J \sum_{i=1}^K \Psi^{(J)}_{Ji} + n_J \sum_{i=1}^L \pi^{(J)}_{JKi} \leqslant a r_2 <\\
	& < \sum_{i=1}^{J} \Sigma_i + n_J \sum_{i=1}^{K+1} \Psi^{(J)}_{Ji} + n_J \sum_{i=1}^{L+1} \pi^{(J)}_{JKi}.
\end{aligned}
\end{equation}

Note how equations \eqref{eq:samplingPopulation} and (\ref{eq:samplingRelation}--\ref{eq:samplingReaction}) are similar, but their implementation in the pseudocode (algorithm \ref{al:reactionSampling}) is different.
This design minimizes sampling errors due to floating point representation, making the first stage exact.

The resulting reaction sampling finds a reaction in exactly the same manner as equation \eqref{eq:reactionGillespie} does.
However, the first and the second stages of this sampling process take $\mathcal{O}(N)$ steps and the third step takes $\mathcal{O}(\alpha)$ steps, resulting in a total time complexity of $\mathcal{O}(N+\alpha)$.
Using \eqref{eq:reactionGillespie} directly requires $\mathcal{O}(M)$ steps\cite{Gillespie1976}, which is $\mathcal{O}(\alpha N^2)$ for densely connected reaction networks.

\begin{figure}[t]
\begin{algorithm}[H]
\caption{Reaction sampling}
\label{al:reactionSampling}
\begin{algorithmic}[1]
\Function {SampleReaction} {$tp$, $r_2$}
   	\State $s_1 \gets 0$, $s_2 \gets 0$
	\While{$ar_2 > s_1$}
    	\State Get a new \texttt{Population} $p_i$ from $tp$'s list
		\State $s_2 \gets s_1$, $s_1 \gets s_1 + \Sigma_i$
	\EndWhile
    \State $ g \gets (ar_2 - s_2) / n_i $
    \While{$ g > 0 $}
    	\State Get a new \texttt{Relation} $\rho_{ij}$ from $p_i$'s list
    	\State $ g \gets g - \Psi^{(i)}_{ij} $
    \EndWhile
	\State $ g \gets g + \Psi^{(i)}_{ij} $
	\While{$ g > 0 $}
    	\State Get a new \texttt{Reaction} $R_{ijk}$ from $\rho_{ij}$'s list
        \State $ g \gets g - \pi^{(i)}_{ijk} $
    \EndWhile
    \Return{$R_{ijk}$}
\EndFunction
\end{algorithmic}
\end{algorithm}
\end{figure}

\subsection{Updating stage}
\label{sec:update}

\begin{figure}[t]
\begin{algorithm}[H]
\caption{Post-sampling data update}
\label{al:dataUpdate}
\begin{algorithmic}[1]
\Function {UpdateExistingPopulations} {$tp$, $R_{IJK}$}
	\ForAll{species $S_i$ participating in $R_{IJK}$}
    	\State Search for a \texttt{Population} of specie $S_i$ in $tp$'s list
        \Statex[2] using its $ID_i$
        \If{$S_i$ has a \texttt{Population} $p_i$ in $tp$'s list}
        	\State $n_i \gets n_i + \sigma_i(R_{IJK})$
            \ForAll{\texttt{RelationAddress}es $RA_{ij}$ in $p_i$'s list}
            	\State Follow the references at $RA_{ij}$ to get 
                \Statex[5] the \texttt{Relation} $\rho_{ji}$ and its owner
                \Statex[5] \texttt{Population} $p_j$
                \State $\Lambda^{(j)}_j \gets \Lambda^{(j)}_j - \Psi^{(j)}_{ji}$
                \State $\Psi^{(j)}_{ji} \gets 0$
                \ForAll{\texttt{Reaction}s $R_{jik}$ in $\rho_{ji}$'s list}
                	\State Recompute $\pi^{(j)}_{jik}$ using \eqref{eq:partialPropensityForReactionTypes}
                    \State $\Psi^{(j)}_{ji} \gets \Psi^{(j)}_{ji} + \pi^{(j)}_{jik}$
                \EndFor
                \State $\Lambda^{(j)}_j \gets \Lambda^{(j)}_j + \Psi^{(j)}_{ji}$
                \State $\Sigma_j \gets n_j \cdot \Lambda^{(j)}_j$
            \EndFor
        \Else
        	\State \Call{AddPopulation}{$tp$, $ID_i$, $\sigma_i(R_{IJK})$}
        \EndIf
    \EndFor
    \ForAll{\texttt{Population}s $p_i$ in $tp$'s list}
    	\If{$n_i == 0$}
        	\State \Call{DeletePopulation}{$tp$, $p_i$}
        \EndIf
    \EndFor
\EndFunction
\end{algorithmic}
\end{algorithm}
\end{figure}

When the reaction to occur $R_{JKL}$ is known, our algorithm proceeds to update the data to reflect the changes in species' populations and propensities (see also algorithm \ref{al:dataUpdate}).
For every specie $S_i$ involved in $R_{JKL}$ we read its $ID_i$ from the array stored at $R_{JKL}$.
We run a sequential search for this specie's \texttt{Population} $p_i$ over the list kept in \texttt{TotalPopulation}.
If the specie's \texttt{Population} is found, its molecular count $n_{i}$ is updated using the stoichiomethic coefficient $\sigma_{i}(R_{JKL})$:
\begin{equation}
	n_{i} \gets n_{i} + \sigma_{i}(R_{JKL}).
\end{equation}
Stoichiometric coefficients are negative for reagents, so their molecular counts may become zero after this step.

After updating the molecular count, we also update all partial and total propensities which depend on it.
From every \texttt{RelationAddress} $RA_{ij}$ in $p_i$'s list we obtain references to a \texttt{Relation} $\rho_{ji}$ in which $S_i$ participates and to the \texttt{Population} $p_j$ of its owner specie.
For $\rho_{ji}$ we recompute all $\pi^{(j)}_{jik}$ and $\Psi^{(j)}_{ji}$ from scratch using formulas \eqref{eq:partialPropensityForReactionTypes} and \eqref{eq:psiDefinition}.
For $p_j$, we update the propensity sums as follows:
\begin{equation}
\begin{aligned}
	&\Lambda^{(j)}_j \gets \Lambda^{(j)}_j - \Psi^{(j)}_{ji},\\
    &\Sigma_j \gets n_j \Lambda^{(j)}_j.
\end{aligned}
\end{equation}

Since each of the species involved in $R_{JKL}$ may be involved in $\mathcal{O}(\alpha N)$ reaction, updating the structure in this way takes a total of $\mathcal{O}(\alpha N)$ operations.

If the specie $S_i$ is a product, its \texttt{Population} may not exist yet.
In this case we add a new \texttt{Population} of $S_i$ using its $ID_i$ as described in section \ref{sec:addition}.
The molecular count of the newly added specie is its stoichiometric coefficient in $R_{JKL}$, $\sigma_{i}(R_{JKL})$.
Additions take $\mathcal{O}(\alpha N)$ operations.

When we're done updating the existing \texttt{Population}s and adding the new ones, we iterate through the list of \texttt{Population}s again and delete the ones with a molecular count of zero as described in section \ref{sec:deletion}. The deletion takes $\mathcal{O}(\alpha N)$.

Finally, we recompute the total propensity of the system $a$ using \eqref{eq:aDefinition}.

\subsection{Summary of EPDM}

EPDM is an SSA which only maintains the data about the species with nonzero molecular count (see algorithm \ref{al:coreEPDM}).
This ensures that the number of tracked molecular species $N$ is as low as possible.

Our algorithm uses a data structure described in section \ref{sec:dataStructure}. The structure holds one entry for each possible reaction, bringing storage requirements of our algorithm to $\mathcal{O}(\alpha N^2)$.

The data structure is initialized by constructing it to be empty, then adding the specie data for every specie initially present in the system. The process is described in section \ref{sec:initialization} and takes $\mathcal{O}(\alpha N^2)$ operations.

Each step of the simulation executes a single reaction. It is composed of a sampling step (see section \ref{sec:sample}) and an updating step (section \ref{sec:update}). Each of these takes $\mathcal{\alpha N}$ operations, so the total number of operations needed to simulate one reaction is also $\mathcal{O}(\alpha N)$.

When some reaction produces any number of molecules of a previously unknown specie, specie data is added as described in section \ref{sec:addition}.
To generate the reactions dynamically, a user-defined function \texttt{reactions()} is used which takes two species and produces a list of reactions between them, complete with rates.

When any specie's molecular count reaches zero, its data is pruned from the structure as described in section \ref{sec:deletion}.

\begin{figure}[t]
\begin{algorithm}[H]
	\caption{EPDM overview}
	\label{al:coreEPDM}
	\begin{algorithmic}[1]
    	\LineComment{Initialization stage, see section \ref{sec:initialization}}
    	\State Create TotalPopulation $tp$ based on initial conditions
        \While{$a>0$ \textbf{and} termination conditions not met}
            \LineComment{Sampling stage, see section \ref{sec:sample}}
        	\State Generate two random numbers $r_1$ and $r_2$
        	\State Compute $\tau$ using \eqref{eq:tauGillespie}
            \State $R\gets$\Call{SampleReaction}{$tp$, $r_2$}
            \LineComment{Updating stage, see section \ref{sec:update}}
            \ForAll{species $S_i$ involved in $R$}
				\State Search $tp$'s list for \texttt{Population} of $S_i$
                \If{a \texttt{Population} $p_i$ of $S_i$ has been found}
                	\State Update $S_i$'s molecular count $n_i$
                    \State Update all partial propensities which
                    \Statex[4] depend on $n_i$
                \ElsIf{no \texttt{Population} of $S_i$ has been found}
                	\State Add a \texttt{Population} of $S_i$ to $tp$'s list
                	\LineComment{details in section \ref{sec:addition}}
                \EndIf
			\EndFor
            \State Find all \texttt{Population}s with $n==0$ in $tp$'s list
            \State Remove all found \texttt{Population}s
            \LineComment{details in section \ref{sec:deletion}}
        \EndWhile
	\end{algorithmic}
\end{algorithm}
\end{figure}

\subsection{Implementation}

We implemented our algorithm as a C++ framework.
The user must define a class \texttt{Specie} with a constructor from a string $ID$ which must save the string into the member variable \texttt{m\_id}.
The class must define a method \texttt{Specie::reactions(Specie)} returning a list of possible \texttt{Reaction}s between the caller \texttt{Specie} and the argument.

After implementing the class \texttt{Specie}, users can simulate the system.
Two stopping criteria are currently available by default: the algorithm can stop either when a certain number of reactions have been executed or a certain simulation time has passed.

A global dictionary with arbitrary parameters loaded from a configuration file is provided for convenience.

The implementation is currently available for Linux and Mac OS X. It is tested with GNU gcc 4.9.3 and GNU make 4.1.

The code is available at \url{https://github.com/abernatskiy/epdm}.

\section{Benchmarks}

We benchmark performance of EPDM against two direct methods: PDM\cite{Ramaswamy2009}\footnote{We took implementation from 
\href{http://mosaic.mpi-cbg.de/pSSALib/pSSAlib.htm}{http://mosaic.mpi-cbg.de/pSSALib/pSSAlib.html}}
and DM\cite{Sanft2011b}\footnote{We took implementation from 
\href{http://sourceforge.net/projects/stochkit/}{http://sourceforge.net/projects/stochkit/}}

Because our model was designed with complex systems in mind we studied performance only for strongly coupled systems.
In both systems every specie can interact with every other specie in a unique way, ensuring that the number of reactions is
\begin{equation}
	M = \frac{N(N+1)}2
\end{equation}
and
\begin{equation}
	\alpha = 1.
\end{equation}
Both models are designed in such a way that the total number of species is preserved throughout the simulation time.

Models below are designed to measure the performance of our method.
They don't fully illustrate the power of the model because they have fixed number molecules, which is necessary to keep for benchmark.
The most striking performance gain is achieved when listing all the species isn't possible in principle or due to computational costs.
For example, in case of realistic polymerization and autocatalysis model used to study prebiotic polymerization~\citeauthor[in press]{Guseva2016} it was possible to  increase the maximum length of simulated polymers from 12 to 25 by employing our algorithm.
Note that limit of 25 wasn't due to restrictions of our algorithm, but due to necessity to calculate minimum energy folding configuration of every chain, which is an NP-complete problem.

\subsection{CPU time of EPDM is linear for a strongly coupled system}
\label{sec:collidingParticles}

\begin{figure}[hbt!]
\centering
\includegraphics[width=\columnwidth]{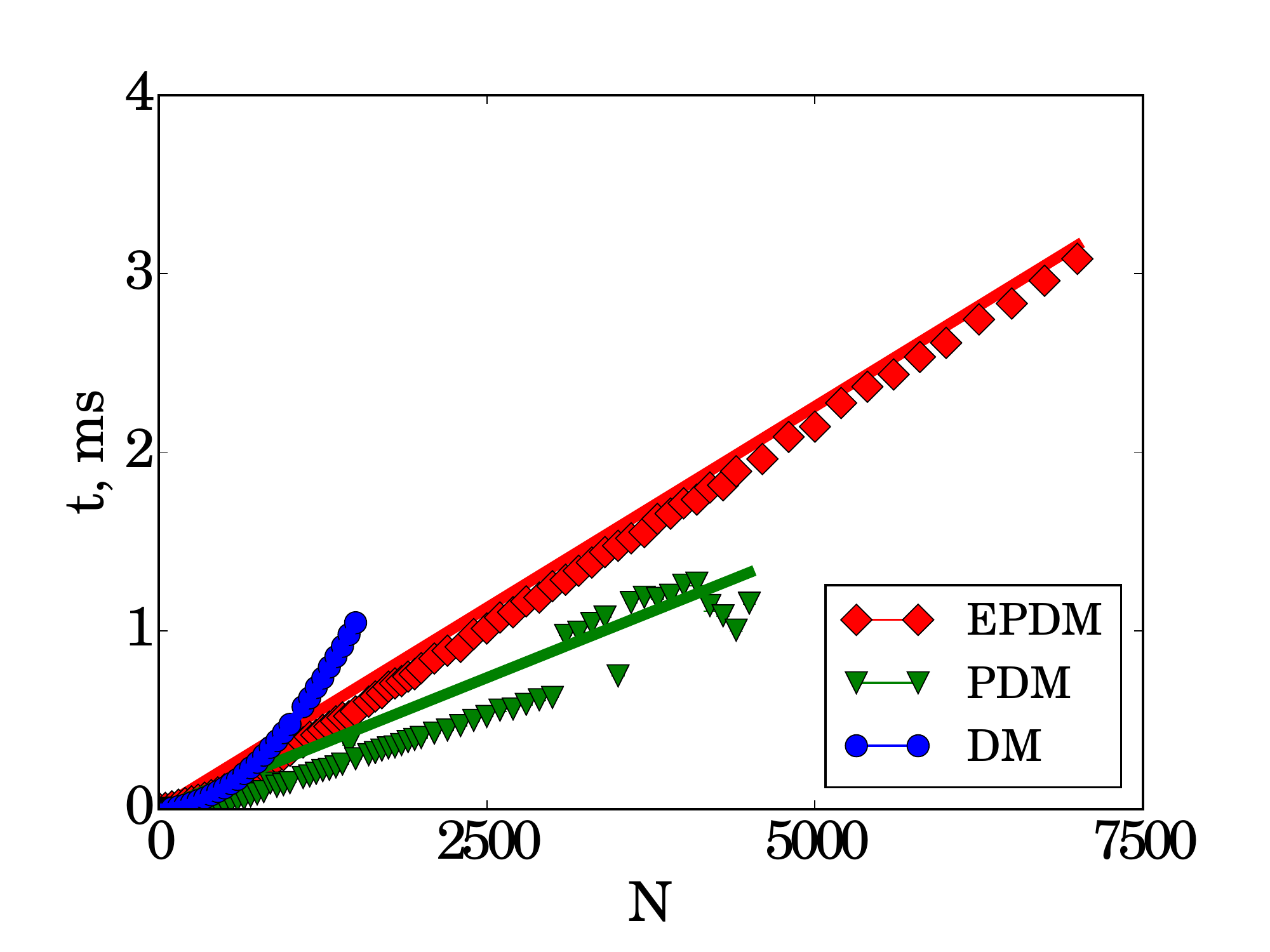}
\caption{CPU time per reaction as a function of the number of species in the system for "colliding particles" model (section \ref{sec:collidingParticles}) for EPDM, PDM and DM.
}
\label{fig:coll_part}
\end{figure}

The first model ("colliding particles") is made to test how our algorithm performs on chemical systems where no new molecules are created and no molecules ever disappear. 
This is a model of a system consisting of colliding particles of $N$ species.
Particles behave like rigid spheres: they collide, and bounce back without internal changes.
All of the particle species are known in advance, none are added or removed over the course of the simulation.  
The system is defined by the following set of equations:
\begin{equation}
  P_i+P_j\xrightarrow{k}P_i+P_j,\quad i,j\in[1,N].
\end{equation}

In our simulations we vary number of species $N$ from $10$ to $7000$.
Every specie has a population of $50$ molecules. 
Collision rate $k$ is fixed at $0.5$ $s^{-1}$. 
Every simulation runs until $5000$ reactions have occurred.
For every value of $N$, CPU time to simulation completion was measured $10$ times and the average time is reported.

Figure \ref{fig:coll_part} shows CPU time it takes per reaction for DM, PDM and EPDM. 
DM is clearly quadratic. 
For any given $N$ PDM outperforms the EPDM, but both are linear in time.
It is important to note that DM and PDM were stopped for a relatively low values of $N$ due to excessive RAM consumption (about $\sim120$GB) by both of the applications, which we suspect was due to implementation issues (in particular, in XML handling libraries).

\subsection{CPU time stays linear when species are actively deleted and created}
\label{sec:particlesWithColor}

The second test checks if algorithm keeps linear time when \texttt{Population}s are added and removed from the system.
The test system ("particles with color") consists of $N$ colliding particles of $N$ types, each of which has an internal property ("color") that is changed in the collision.
The following equation defines the system:
\begin{equation}
P_i^{\alpha}+P_j^{\beta}\xrightarrow{k}P_i^{\alpha+1}+P_j^{\beta+1},\quad i,j\in[1,N],\,
\alpha,\beta\in[1,\Omega]
\end{equation}
Indexes $i,j$ enumerate particle types and run from $1$ to $N$.
Indexes $\alpha, \beta$ enumerate colors of particles; particle can have one of $\Omega$ colors. 
During the collision color index of a participating particle goes up $P_i^{\alpha}\to P_i^{\alpha+1}$, until it reaches maximal index $\Omega$, after which it drops back to $0$: $P_i^{\Omega}\to P_i^{0}$.

Since every combination of a particle type and a color is considered a separate specie, every reaction causes two species to go extinct and their \texttt{Populations} to be deleted. 
It also adds two new species, which requires adding two new \texttt{Populations}. 
Thus, the total number of species simultaneously present in the system is maintained at exactly the same level.
Every specie is represented in the system by a single molecule.

\begin{figure}[hbt!]
\centering
\includegraphics[width=\columnwidth]{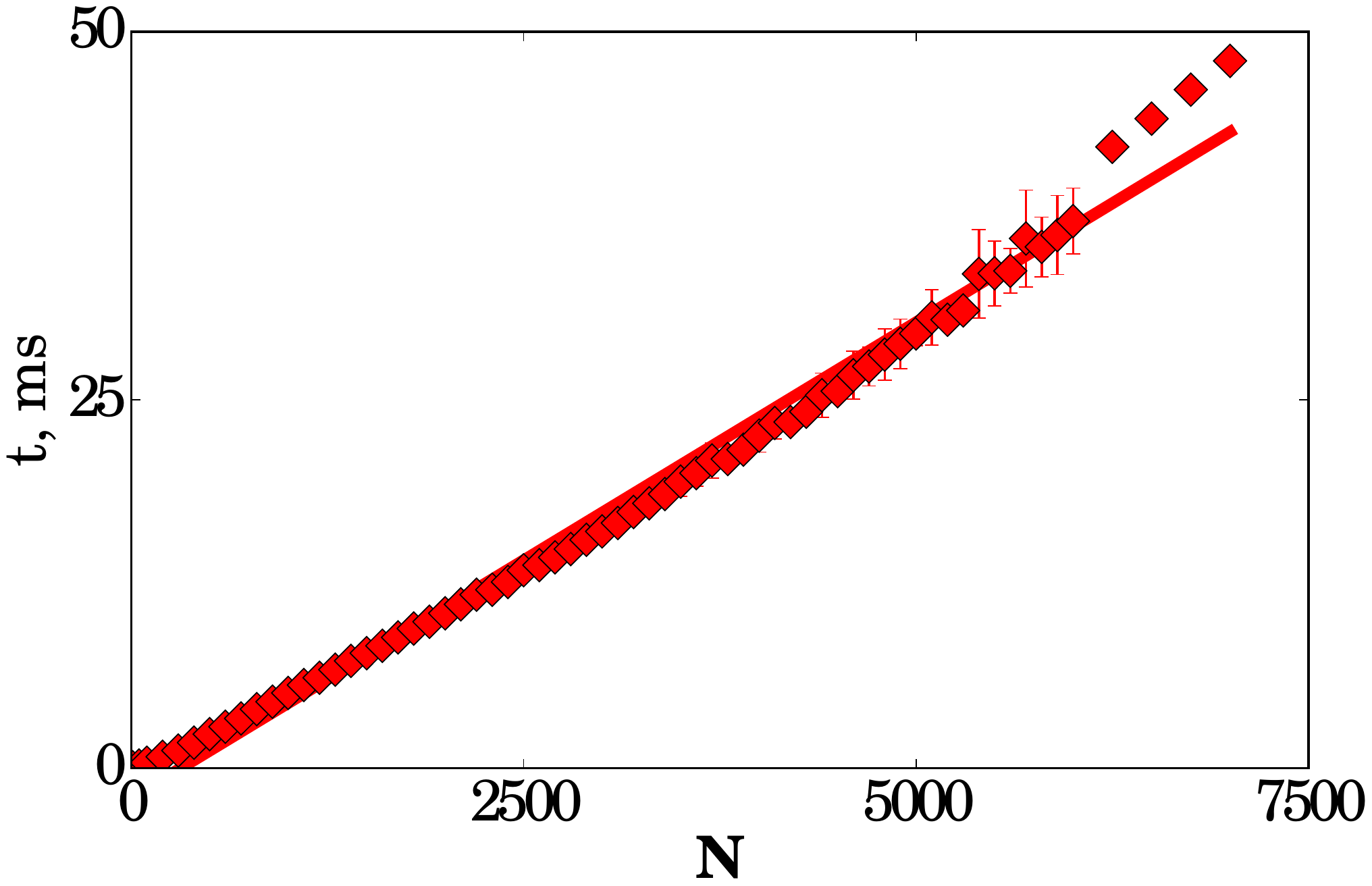}
\caption{CPU time per reaction for EPDM simulation of "particles with color" (see section \ref{sec:particlesWithColor}).}
\label{fig:coll_part_del}
\end{figure}

To run such a simulation in DM and PDM frameworks we had to enumerate all $N\Omega$ of the possible species. 
This slows down the simulation enough to make the comparison impossible beyond a small number of species.
The figure \ref{fig:coll_part_del} shows how CPU time per reaction depends on the number of species in EPDM.

\section{Conclusion and Discussion}
Stochastic simulations are actively used in molecular and systems biology.
The bigger and more complex the system, the more important the performance of the simulation algorithm becomes.
It is also more burdensome or even impossible to list all the species and reactions for complex systems.
We introduced a general purpose, exact stochastic simulation algorithm which allows to avoid listing all the possible species and reactions by defining the general rules governing the system instead. 
Built within the partial propensity framework \cite{Ramaswamy2009,Ramaswamy2010}, our algorithm achieves linear time complexity in the number of molecular species.

The algorithm has its limitations. 
First, it is limited to reactions of maximum of two reactants. 
In the chemical system that doesn't present much of an issue because reactions with three molecules are significantly rarer then binary reactions and more complex reactions can be represented as sequences of binary reactions.
Second, it cannot simulate spatially nonhomogeneous systems.
However, as long as all reactions involve no more than two reactants and the observation from section \ref{sec:gillespie} holds, it is possible to simulate any system for which the set of reactions between any two species is known.

Benchmarks suggest that our algorithm is slower than PDM by a constant factor. 
Therefore, one should use PDM when it is unlikely that a significant proportion of species will have a molecular count of zero.

Our results suggest that in complex systems (such as polymerization reactions and networks of intra-cellular reactions) and in the case of emergent phenomena studies (e.g. origins of life) EPDM can significantly improve the performance of the stochastic simulations.
The software implementation of the algorithm is available as an open source public repository \url{https://github.com/abernatskiy/epdm}.

\section{Acknowledgments}
Authors appreciate support from the NSF grant MCB1344230.

\bibliographystyle{apalike}
\bibliography{library}

\end{document}